\documentclass{aa}

\usepackage{graphicx}
\usepackage{txfonts}
\usepackage{hyperref}
\usepackage{bm}
\usepackage{xcolor}

\begin{document}

   \title{Derivation of generalized Kappa distribution from scaling properties of solar wind magnetic field fluctuations at kinetic scales}

   \titlerunning{Derivation of generalized Kappa distribution from scaling properties of solar wind}


\author{Daniele Belardinelli\inst{1}\thanks{These authors contributed equally to this work.}
\and Simone Benella\inst{2}\textsuperscript{$\star$}
\and Mirko Stumpo\inst{2}
\and Giuseppe Consolini\inst{2}
}

\institute{Department of Physics, University of Rome Tor Vergata,
Via della Ricerca Scientifica 1,
Rome, 00133, Italy
\and
INAF - Institute for Space Astrophysics and Planetology,
Via del Fosso del Cavaliere 100,
Rome, 00133, Italy\\
}


 
\abstract
{Kinetic scale dynamics in weakly-collisional space plasmas usually exhibits a self-similar statistics of magnetic field fluctuations which implies the existence of an invariant probability density function (master curve).}
{We provide an analytical derivation of the master curve by assuming that perpendicular fluctuations can be modeled through a scale-dependent Langevin equation.}
{In our model, magnetic field fluctuations are the stochastic variable and their scale-to-scale evolution is assumed to be a Langevin process. We propose a formal derivation of the master curve describing the statistics of the fluctuations at kinetic scales. Model predictions are tested on independent data samples of fast solar wind measured near the Sun by Parker Solar Probe and near the Earth by Cluster.}
{The master curve is a generalization of the Kappa distribution with two parameters: one regulating the tails and the other one controlling the asymmetry. Model predictions match the spacecraft observations up to 5$\sigma$ and even beyond in the case of perpendicular magnetic field fluctuations.}
{}


\maketitle
%

\section{Introduction}
\nolinenumbers

In the context of space and astrophysical plasmas, which encompasses environments like solar and stellar winds, planetary magnetospheres, and the interstellar medium, the dynamic interplay of energy fluctuations, related to various physical processes, unfolds across an extensive range of spatial and temporal scales. Diverse structures, including current sheets, plasmoids, and vortices, emerge as a consequence \citep{Coleman1968,Higdon1984,Zhuravleva2014}. Large scale velocity and magnetic field fluctuations, where plasma behaves like a single fluid, show the typical hallmarks of fully developed turbulence \citep{Belcher1971,biskamp_2003}. Conversely, at scales smaller than characteristic ion length (kinetic scales) where the motion of ions and electrons decouples, the interaction between different kinetic processes yields fluctuations with distinct spectral and statistical features \citep{Cranmer2003,Bale2005,chen2013,bruno16}. Thanks to space missions which are providing solar wind observations of increasingly higher resolution over the last decades, it was possible to investigate statistical properties of magnetic field fluctuations, from the fluid to the ion-kinetic regime \citep{chasapis2017,chhiber2021subproton}.

In the attempt to understand the dynamics at kinetic scales in the solar wind, statistical methods have emerged as indispensable tools, as they are closely related to fundamental physical quantities, such as the energy transferred at different scales in the system, and are subject of theoretical predictions \citep{bruno16}. In turbulence, for example, a careful examination of the common statistical properties shared by flows across diverse systems, viz., from fluids to both laboratory and space plasmas, allowed to introduce the concept of universality \citep{Lvov1998,frisch1995,Arneodo2008}. The idea of universality consists in the emergence of common statistical properties (e.g., scaling exponents, spectral slope, etc.) in a large set of independent data samples, indicating that statistical laws emerge regardless of the specific processes underlying energy injection/dissipation. In the case of space plasma turbulence, e.g., the emergence of a power law in the power spectra of velocity and magnetic fields, $E(f)\sim f^{-\alpha}$ as a function of frequency, is ubiquitous in spacecraft data, with $\alpha$ very close to 5/3 or 3/2 at inertial scales (i.e., at scales larger than the ion inertial length and smaller than the correlation length), in agreement with the predictions by \cite{Kolmogorov1941} and \cite{kraichnan65}, respectively. 
Conversely, the scaling exponents of the spectral density routinely observed at kinetic scales do not show universal values, thus being strongly dependent on the considered data sample \citep{sahraoui2013scaling,carbone2022high,Sahraoui2023}. This suggests that at scales smaller than the ion inertial length there is a transition towards a dynamical regime formed by processes of different nature with respect to the magnetohydrodynamic turbulence. The search for universal properties in this range would be important in order to test if even kinetic-scale physics is characterized by the emergence of universal dynamics. However, universality is to be sought elsewhere, given that spectral exponents do not exhibit this property. 
In contrast to what is expected for a turbulent plasma, one of the striking features of kinetic scale dynamics highlighted from several data samples in both spacecraft measurements and numerical simulations, is the self-similar statistics of magnetic field fluctuations, which implies the existence of a scale invariant probability density function --- master curve --- in this range of scales \citep[see, e.g.,][]{Consolini2005,kiyani2009global,osman2015,leonardis2016}. Theoretical frameworks able to describe such properties and provide mathematical models for probability distribution functions (PDFs), scaling exponents, and so on, are therefore of primary importance in the search for universal behaviors at small scales. Among these, methods based on stochastic process theory are very effective in capturing and predicting statistical properties of fluctuating time series \citep[e.g, see the book][]{tabar2019analysis}.

Descriptions of turbulent and kinetic dynamics based on stochastic processes have recently emerged in the field of astrophysical plasmas such as solar photosphere \citep{Ramos2014,Gorobets2016}, solar wind \citep{strumik2008testing,carbone2021on,Bian2022,benella2022markovian,carbone2022high,Bian2024a,Bian2024b} and Earth magnetosheath \citep{macek2022mms,Chiappetta2023,Macek2023MNRAS,Wojcik2024}. Many of these models are designed to predict PDFs and infer stochastic equations governing the fluctuations at different scales, thereby extending earlier findings from fluid turbulence \citep[e.g., see][]{friedrich1997,renner2001experimental} to space plasmas. In the case of plasmas, both velocity and magnetic field increments at different scales can be envisioned as Langevin processes where the physics of evolution and interaction of structures is enclosed within stochastic diffusion terms. The statistics of these processes offer a description which is equivalent to scaling models, provided that the drift and diffusion terms are chosen appropriately \citep{Nickelsen2017,Friedrich2020}.
In this context, the aim of this work is to use the generalized Langevin equation to formally derive a master curve representative of the magnetic field fluctuations observed in space plasmas at kinetic scales. The derivation of an analytical expression for this distribution may be important in the search for new universal properties of kinetic scale fluctuations. As we will show, the PDF can be expressed as a Kappa distribution. Kappa distributions play a fundamental role in the field of space plasmas \citep{livadiotis2017kappa,Lazar2021}. Empirically introduced by \cite{Olbert1968} and \cite{Vasyliunas1968} to model electron fluxes observed by early space missions, these distributions have led to significant advancements in understanding and modeling space plasma dynamics, being related to various fundamental mechanisms such as particle acceleration \citep{Zank2006,Fisk2014,Bian2014}, turbulence \citep{Leubner2005a,Yoon2012,Yoon2014,Gravanis2019,Yoon2020}, system for which the temperature (or other relevant thermodynamic quantities) varying according to a certain distribution rather than being constant \citep[i.e., superstatistics;][]{Beck2003,Schwadron2010,Livadiotis2019,Gravanis2020}, pick-up ion cooling \citep{Livadiotis2012,Livadiotis2024}, and many others. Indeed, Kappa distributions are particularly effective in describing different plasma populations in the heliosphere \citep{Collier1996,Maksimovic1997,Pierrard1999,Mann2002,Marsch2006,Zouganelis2008,Livadiotis2010,Pavlos2016,livadiotis2017kappa,Lazar2021}. 

Focusing on solar wind magnetic field fluctuations at kinetic scales, we aim to derive the Kappa distribution from the scaling properties of the magnetic field. To this purpose, in Section \ref{sec:phe} we summarize some relevant scaling theories of turbulence and to highlight their connection with the Langevin description; in Section \ref{sec:met} we illustrate the solution of the associated stationary Fokker-Planck equation (FPE) leading us to the Kappa distribution; in Section \ref{sec:res} we test predictions against solar wind data, and in Section \ref{sec:con} we draw our conclusions.

\section{A stochastic model for magnetic field increments}
\label{sec:phe}
In magnetohydrodynamic turbulence, it is a common practice to use velocity/magnetic field increments as a proxy, on average, of the amount of energy contained at a given scale $r$ into the system. As the energy is transferred from the injection scales towards dissipative scales, the amount of energy, and thus the level of fluctuations, decreases for decreasing $r$. Statistical analysis involves high-order moments of increments $X_r$ of a field $F$ defined as:
\begin{equation}
    X_r := F(x + r) - F(x),
\end{equation}
at a scale separation $r$. If these increments exhibit global scale invariance, large structures divide into smaller and smaller structures. If this process is homogeneous, i.e., each structure decays in the same manner, the field $F$ is said to be globally self-similar or globally scale invariant. Formally, the global invariance can be expressed as
\begin{equation}
    r\rightarrow\lambda r\Rightarrow X_{\lambda,r} = \lambda^h X_r,
\end{equation}
where $h$ is the scaling exponent and $\lambda \in \mathbb{R}+$ is a dilation parameter. This results in the power-law behavior of the individual realization $X_r$ of the process and of the structure functions, $S_q(r)\sim r^{\zeta_q}$, with $\zeta_q$ as the scaling exponents. Self-similarity hypothesis was part of the first theory proposed by \cite{Kolmogorov1941}. In particular, he proposed the linear relation for the scaling exponents $\zeta_q = q/3$, where $q$ indicates the order of the moment. In this context, $\zeta_q$ signifies how increments at different scales contribute to the overall statistical behavior of turbulence.
Given the fluctuation level $X_r$ at the scale $r$, a simple choice accounting for subtracting energy in a self-similar way is represented by the damping equation:
\begin{equation}
    dX_r=M(X_r,r)dr=m(r)X_r\,dr,
\end{equation}
where the drift term depends on the scale $r$ and the process $X_r$. Assuming that $m(r)\sim h\,r^{-1}$, the solution of the damping equation is a power law:
\begin{equation}
    X_r=X_0\bigl(r/r_0\bigr)^h.
    \label{Eq:damp}
\end{equation}
Despite its simplicity, which makes it unsuitable for describing a typical turbulent dynamics, it has been recently pointed out how the damping trend may constitutes a reasonable zeroth-order approximation of kinetic scale fluctuations in the solar wind \citep{Benella2023}.
Nevertheless, Equation (\ref{Eq:damp}) produces deterministic trajectories which behave as a power laws with a unique exponent $h$, whereas the typical dynamics of the magnetohydrodynamic regime is more rich and chaotic, i.e., it typically exhibits a set of $h$ depending on space and on the separation scale \citep{Parisi85,frisch1995}. 
As a result, the scaling exponent $\zeta_q$ of the structure functions turns out to be a nonlinear function of the order $q$. This statistical signature is known as anomalous scaling, which is directly related to intermittency and which has been captured through a plethora of statistical models, e.g., log-normal scaling, also known as the refined similarity hypothesis by Kolmogorov in 1962 \citep[K62;][]{Kolmogorov1962,Obukov1962}, log-Poisson model \citep{Dubrulle1994}, She-Leveque model \citep{She1994}, random cascade models \citep{Castaing1995}, Yakhot model \citep{Yakhot1998}, and so forth. In terms of the energy cascade mechanism, intermittency is related to non-homogeneity of the energy transfer/dissipation rate, i.e., the efficiency of the energy cascade depends on space and time.

In the context of stochastic description of the changing fluctuations across the scales, we can introduce a stochastic term characterized by pure multiplicative noise in Equation (\ref{Eq:damp}), thus ending up with a real Itô stochastic process defined through a generalized Langevin equation
\begin{multline}
    dX_r=M(X_r,r)dr + \sqrt{2D(X_r,r)}dW_r\\
    =m(r)\,X_r\,dr+\alpha X_r\,dW_r,
    \label{Eq:langevin}
\end{multline}
where $W_r$ is a standard Wiener process with zero mean and unit variance and $D(X_r,r)=D(X_r)=\alpha X_r^2/2$ is the diffusion term\footnote{We implicitly assumed $X_r\geq0$ for the sake of simplicity.}. The formal solution of Equation (\ref{Eq:langevin}) is:
\begin{equation}
    X_r=X_0\bigl(r/r_0\bigr)^h\exp\bigl(\alpha W_r-\tfrac{1}{2}\alpha^2r\bigr).
\end{equation}
Therefore, at individual level, we ultimately have a stochastic variable driven by the Wiener process $W_r$. Moreover, since $W_r$ is normally distributed, Equation (\ref{Eq:langevin}) implies that $X_r$ must follows a log-normal distribution, thus resembling, for instance, the K62 correction to scaling \citep{Kolmogorov1962,Obukov1962}. From a stochastic process perspective, the log-normal scaling represents the easiest way to introduce stochasticity in the fluctuations and thus in the transfer of energy at the individual level \citep{Nickelsen2017,fuchs2022}. Nevertheless, a general second order polynomial function can introduce interesting and more realistic features --- viz., the introduction of an additive noise over the multiplicative one --- resembling empirical observations at the cost of the simplicity of providing a formal solution to the stochastic differential equation with respect to the log-normal hypothesis \cite[see][for further details]{reinke2018on}. For sake of generality, we thus introduce a first-order polynomial $L$ and a second-order polynomial $Q$, viz.,
\begin{align}%
    L(X,r) &:= L_0(r) + L_1(r)X,\label{Eq:L} \\
    Q(X,r) &:= Q_0(r) + Q_1(r)X + Q_2(r)X^2, \label{Eq:Q}
\end{align}%
where $Q_0>0$ and $4Q_0Q_2>Q_1^2$, and we will assume, when explicit computation is needed, that $M\equiv L$ and $D\equiv Q$. 

\section{Analytical derivation of the master curve}
\label{sec:met}
In the case of the Langevin process, the probability density $\rho_r(X)$ of finding $X_r$ in the state $X$ satisfies the FPE, which is thus the master equation of the process \citep{risken1996fokker}
\begin{equation}
    \frac{\partial\rho_r(X)}{\partial r} = \frac{\partial}{\partial X}\biggl[\frac{\partial}{\partial X}(D(X,r)\rho_r(X)) - M(X,r)\rho_r(X)\biggr].
\end{equation}
If we consider the standardized process
\begin{equation}
    x_r := \frac{X_r - \mu_r}{\sigma_r}, \label{Eq:x_r}
\end{equation}
where $\mu_r$ and $\sigma_r$ are the scale dependent mean value and standard deviation, the probability density $p_r(x)$ of finding $x_r$ at $x$ is given by
\begin{equation}
    p_r(x) = \sigma_r\rho_r(\mu_r+\sigma_rx)
\end{equation}
and satisfies the FPE with drift and diffusion terms
\begin{align}
    &m(x,r) := \frac{M(\mu_r+\sigma_r x,r) - \frac{d\mu_r}{dr} - \frac{d\sigma_r}{dr}x}{\sigma_r}, \\
    &d(x,r) := \frac{D(\mu_r+\sigma_r x,r)}{\sigma_r^2}\label{eq:diffnorm}.
\end{align}
By choosing $M\equiv L$ and $D\equiv Q$, we get $m\equiv l$ and $d\equiv q$, where
\begin{align}
    &l(x,r) := - [q_0(r) + q_2(r)]x,\label{Eq:l}\\
    &q(x,r) := q_0(r) + q_1(r)x + q_2(r)x^2. \label{Eq:q}
\end{align}
The set of parameters $\{q_0,q_1,q_2\}$ refers to the drift and diffusion terms of the standardized variable and can be expressed in terms of the parameters $\{Q_0,Q_1,Q_2\}$ of Equation \eqref{Eq:Q} as:
\begin{align}%
    q_0(r) &:= \frac{Q(\mu_r,r)}{\sigma_r^2}, \\q_1(r) &:= \frac{Q_1(r) + 2Q_2(r)\mu_r}{\sigma_r}, \\q_2(r) &:= Q_2(r). \label{Eq:q012}
\end{align}
\begin{figure}
    \centering
    \includegraphics[width=0.5\textwidth]{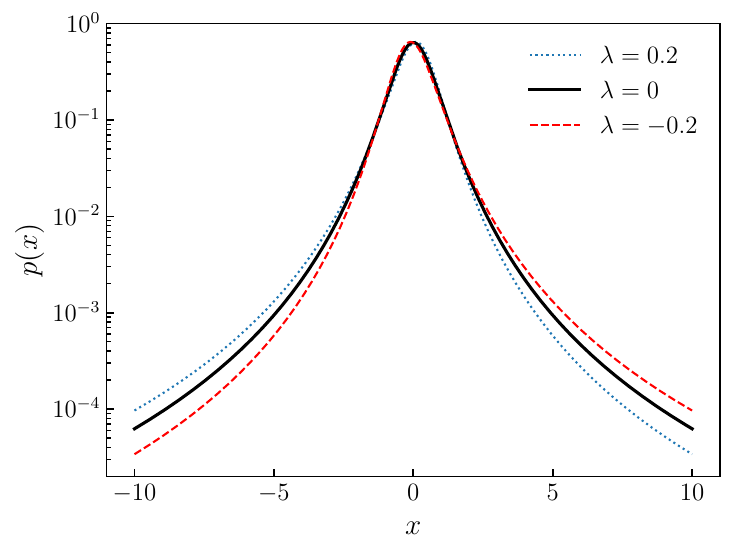}
    \caption{Example of stationary distribution \eqref{Eq:genK} with $\kappa=1$ for different values of $\lambda$. Dashed line indicates $\lambda=-0.2$, dotted line $\lambda=0.2$, and solid line shows the Kappa distribution \eqref{Eq:kappa}.}
    \label{fig:example}
\end{figure}

The evolution of the stochastic process \eqref{Eq:langevin} as a function of the scales, hence, implies a FPE also for the standardized variable $x_r$. This equation, with linear drift and quadratic diffusion, cannot be solved analytically without further hypotheses. Since the statistics of kinetic scales in the solar wind is observed to be monofractal, this implies that the probability density functions can be recast on a master curve by simply rescaling the variable with the standard deviation $\sigma_r$ \citep{Consolini2005,kiyani2009global,osman2015,benella2022markovian}. In this spirit, we assume the stationary condition of the standardized process $(x_r)_{r\geq0}$, i.e., $\partial p_r/\partial r\equiv0$, and we obtain the following analytical solution for the FPE:
\begin{equation}
    p_{\kappa,\lambda}(x) = N_{\kappa,\lambda}\biggl[1 + \frac{1}{\kappa}\frac{(x - \lambda)^2}{x_{\kappa,\lambda}^2}\biggr]^{-\kappa-1}\exp\biggl(- \frac{2\lambda\sqrt{\kappa}}{x_{\kappa,\lambda}}\tan^{-1}\biggl(\frac{x - \lambda}{\sqrt{\kappa}x_{\kappa,\lambda}}\biggr)\biggr),
    \label{Eq:genK}
\end{equation}
where
\begin{align}
    &x_{\kappa,\lambda} := \sqrt{2 - \frac{1 + \lambda^2}{\kappa}},
    \\&N_{\kappa,\lambda} := \frac{2^{2\kappa}\Gamma\left(\kappa + 1 + i\frac{\lambda\sqrt{\kappa}}{x_{\kappa,\lambda}}\right)\Gamma\left(\kappa + 1 - i\frac{\lambda\sqrt{\kappa}}{x_{\kappa,\lambda}}\right)}{\pi\sqrt{\kappa}x_{\kappa,\lambda}\Gamma(2\kappa + 1)},
\end{align}
with $i$ being the imaginary unit and $\Gamma$ the Euler gamma function, and
\begin{align}
    &\lambda = - \frac{q_1(r)}{2q_2(r)}, &&\kappa = \frac{1}{2} + \frac{q_0(r)}{2q_2(r)}.\label{eq:kappalam}
\end{align}
For a detailed formal derivation of these equations we refer to the Appendix \ref{app:a}. Notice that $2\kappa>1+\lambda^2$ by construction. In particular, for $\lambda=0$ we get the familiar Kappa distribution:
\begin{equation}
    p_{\kappa,0}(x) = \frac{\Gamma(\kappa + 1)}{\sqrt{\pi(2\kappa - 1)}\Gamma(\kappa + \frac{1}{2})}\biggl(1 + \frac{1}{\kappa}\frac{x^2}{x_{\kappa,0}^2}\biggr)^{-\kappa-1}.
    \label{Eq:kappa}
\end{equation}
The obtained stationary distribution (\ref{Eq:kappa}) corresponds to the Kappa distribution, and for this reason we refer to Equation \eqref{Eq:genK} as generalized Kappa distribution. We emphasize that the obtained Kappa distribution represents the probability measure of one-dimensional stochastic process $x_r$, and should not be confused with the conventional Kappa distribution in the velocity space defined in three dimensions \citep{Olbert1968,Vasyliunas1968,Lazar2021}. The distribution \eqref{Eq:genK} depends on two parameters: the shape parameter $\kappa$, which regulates the weight of the tails of the distribution, e.g., tending towards a Gaussian for $\kappa\to\infty$ in Equation \eqref{Eq:kappa}, and the symmetry parameter $\lambda$, representing an important and innovative generalization of a Kappa distribution, allowing the distribution to be asymmetric with respect to zero. A sketch of the stationary distribution \eqref{Eq:genK} for three different values of the parameter $\lambda=-0.2,0,0.2$ and $\kappa=1$ is illustrated in Figure \ref{fig:example}. As long as the parameter $\lambda$ is equal to zero, the distribution is symmetric. Conversely, when this parameter is positive/negative one of the tail of the distribution tends to be higher than the other, thus producing non-vanishing odd moments (e.g., non vanishing skewness). Concerning first- and second-order moments, we have
\begin{align}
    &\int_{-\infty}^{+\infty} x p_{\kappa,\lambda}(x)dx = 0, &&\int_{-\infty}^{+\infty} x^2 p_{\kappa,\lambda}(x)dx = 1,
\end{align}
regardless the values of $\kappa$ and $\lambda$ (provided $2\kappa>1+\lambda^2$), as it must be for the random variable $x_r$ in Eq. \ref{Eq:x_r}, by construction. 
\begin{figure*}
    \centering
    \includegraphics[width=\textwidth]{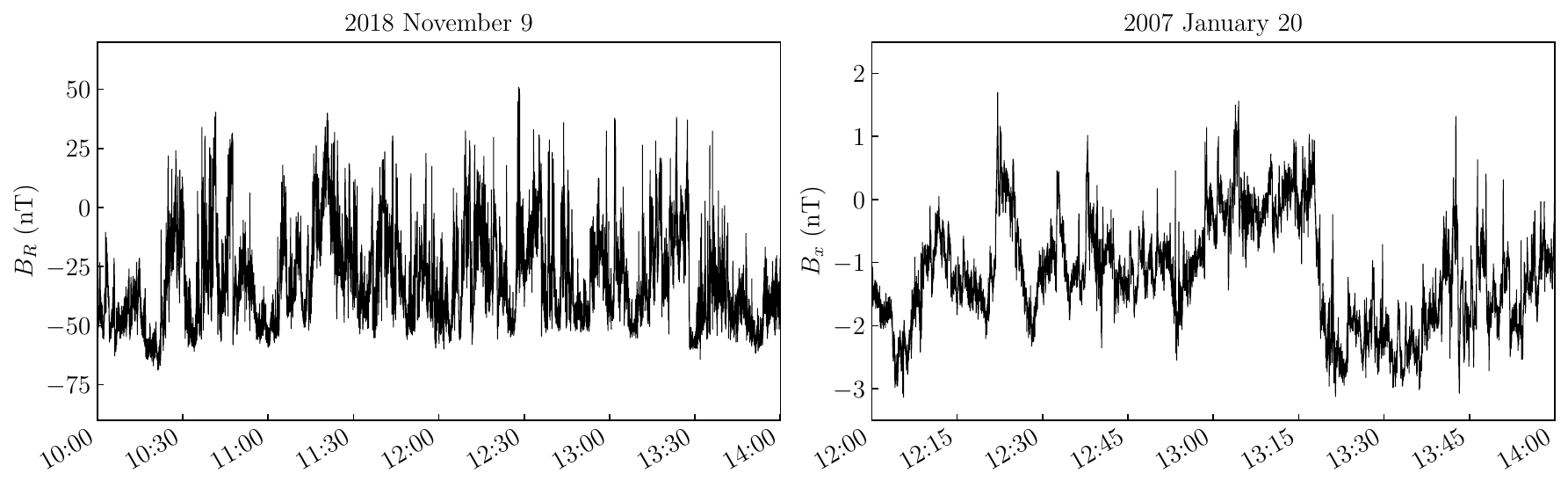}
    \caption{Radial component of the magnetic field $B_R$ observed by PSP during the first perihelion (left), and Magnetic field component $B_x$ in the GSE coordinate system observed by the C3 spacecraft (right).}
    \label{fig:data}
\end{figure*}
\begin{figure*}
    \centering
    \includegraphics[width=\textwidth]{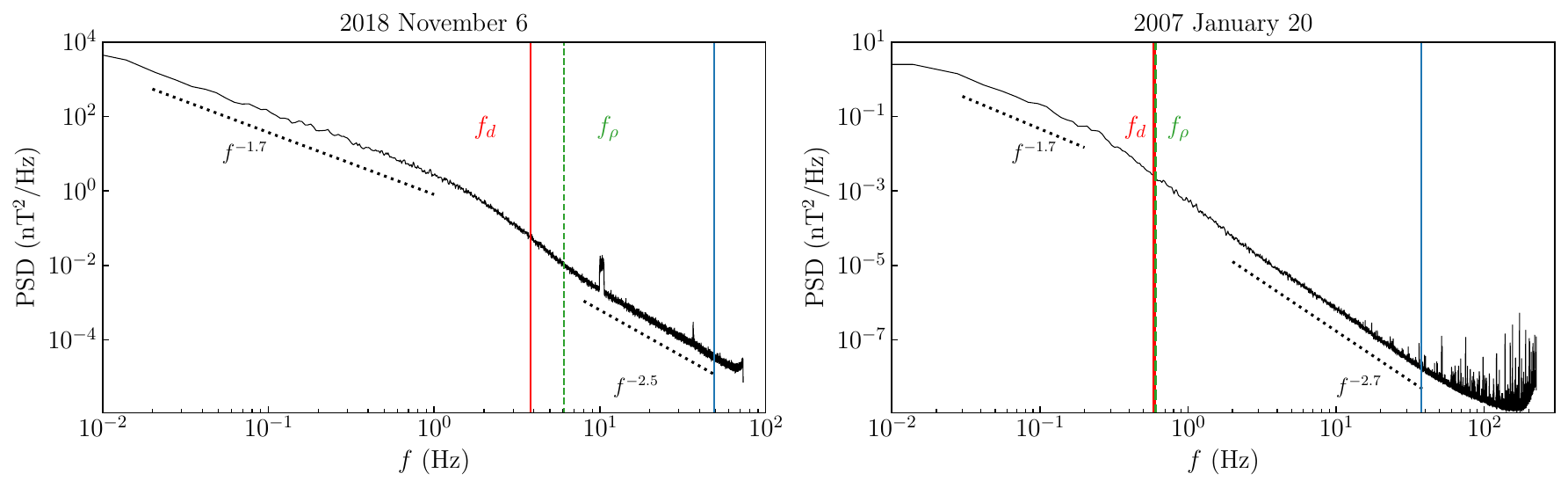}
    \caption{Power spectral density (PSD) of the 2018 November 9 PSP data interval (left) and of the 2007 January 20 C3 data sample (right). Vertical lines indicate frequencies associated with the ion inertial length (red) and the ion gyroradius (green). Blue lines indicate the smallest scales considered in the analysis. The typical slopes in the inertial and kinetic ranges are reported for reference (dotted lines).}
    \label{fig:psd}
\end{figure*}
\begin{table*}
\caption[]{Bulk solar-wind parameters for the data samples used in the analysis.}
\label{tab:1}
\centering          
\begin{tabular}{@{\extracolsep{20mm}}ccc}
\hline
\noalign{\smallskip}
 & Parker Solar Probe & Cluster 3\\
 & 2018 Nov 09, 10:00 -- 14:00 UT & 2007 Jan 20, 12:00 -- 14:00 UT\\
\noalign{\smallskip}
\hline
\noalign{\smallskip}
$B$ (nT) & 56 & 3.7\\
$B_{rms}$ (nT) & 44 & 1.8\\
$B_{\text{max}}$ (nT) & 73 & 4.3\\
$V$ (km s$^{-1}$) & 506 & 600\\
$V_A$ (km s$^{-1}$) & 113 & 53\\
$n$ (cm$^{-3}$) & 117 & 2\\
$T$ ($10^5$ K) & 6.0 & 3.5\\
$d_i$ (km) & 66 & 161\\
$\rho_i$ (km) & 13 & 152\\
$\beta$ & 0.8 & 1.8\\
\noalign{\smallskip}
\hline            
\end{tabular}
\end{table*}

Hence, to sum up, if we consider a stochastic process $X_r$ of the Langevin type defined by the drift and diffusion terms of Equations \eqref{Eq:L} and \eqref{Eq:Q}, if the density function $p_r(x)$ of the standardized process $x_r$ is stationary, the solution is the generalized Kappa distribution \eqref{Eq:genK}.

\section{Results}
\label{sec:res}

In order to test these analytical results on spacecraft observations, we define the longitudinal increment of the interplanetary magnetic field as $b_r:=[\bm{B}(\bm{x}+\bm{r})-\bm{B}(\bm{x})]\cdot\hat{\bm{r}}$, where $\bm{B}$ is the magnetic field vector and $\bm{r}$ is the separation vector between the two measurements. This quantity is our stochastic variable $b_r\equiv X_r$ and the variation of $b_r$ as a function of the scale separations $r$ defines a stochastic process. We consider two independent data samples of fast solar wind streams at different heliocentric distances as a test for the model introduced in the previous section:
\begin{itemize}
    \item The first is an Alfvénic fast wind data interval gathered by the FIELD suite \citep{Bale2016} on board Parker Solar Probe (PSP) during its first perihelion between 02:00 UT and 04:00 UT on 2018 November 9, at an heliocentric distance of approximately 0.17 astronomical units (au). We used the SCaM magnetic field data product, obtained by merging observations gathererd by the fluxgate and the search-coil magnetometers and available with a time resolution of 146.5 samples/s \citep{bowen2020merged}.
    \item The second data sample is measured during a fast wind stream by the Cluster 3 (C3) spacecraft in the near-Earth environment, $\sim1$ au, between 12:00 UT and 14:00 UT on 2007 January 20 \citep{Yordanova2015,alberti2019multifractal}. High-cadence magnetic field data are obtained by the marged measurements of FGM \citep{Balogh1997} and STAFF \citep{Cornilleau1997} instruments, reaching a time resolution of 450 samples/s.
\end{itemize}
More details about average bulk plasma parameters related to the data samples selected for this study are summarized in Table \ref{tab:1}.
\begin{figure}
    \centering
    \includegraphics[width=0.5\textwidth]{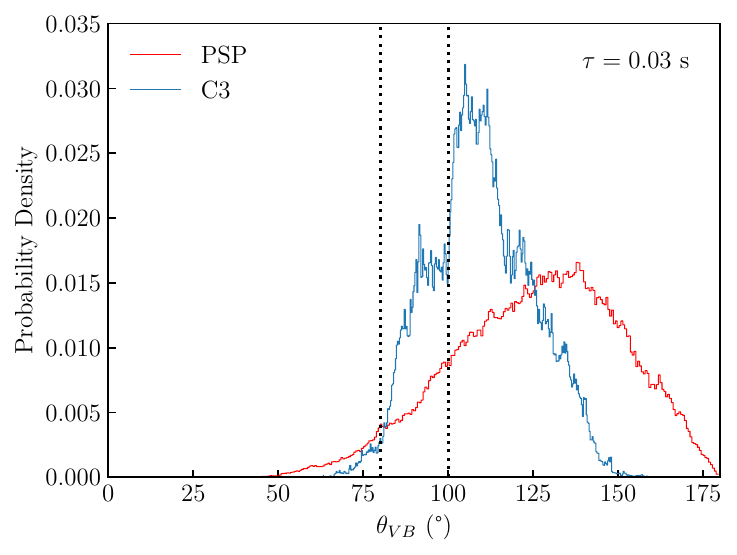}
    \caption{Histograms of the angle $\theta_{VB}$ between the local mean magnetic field and the sampling direction for PSP and C3 data samples. The statistics have been calculated at scale $\tau=0.03$ s and the vertical lines delineate the region $80^\circ<\theta_{VB}<100^\circ$ around the perpendicular direction.}
    \label{fig:theta}
\end{figure}
\begin{figure*}
    \centering
    \includegraphics[width=\textwidth]{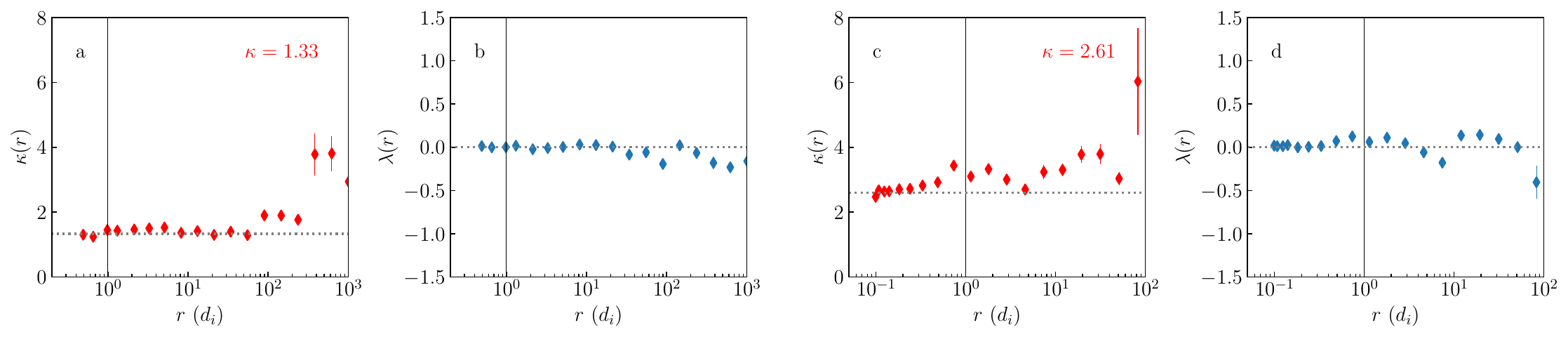}
    \caption{PDF parameters as a function of the scale separation $r$. a) Trend of the parameter $\kappa$ for the PSP data sample. The limit value of $\kappa=1.33$ is highlighted by the horizontal line. b) Trend of the parameter $\lambda$ for the PSP data sample. The value $\lambda=0$ is indicated by the horizontal line. a) Trend of the parameter $\kappa$ for the C3 data sample. The limit value of $\kappa=2.61$ is highlighted by the horizontal line. a) Trend of the parameter $\lambda$ for the C3 data sample. The value $\lambda=0$ is indicated by the horizontal line. Vertical lines in all the panels mark the ion inertial length.}
    \label{fig:kl}
\end{figure*}
\begin{figure}
    \centering
    \includegraphics[width=0.5\textwidth]{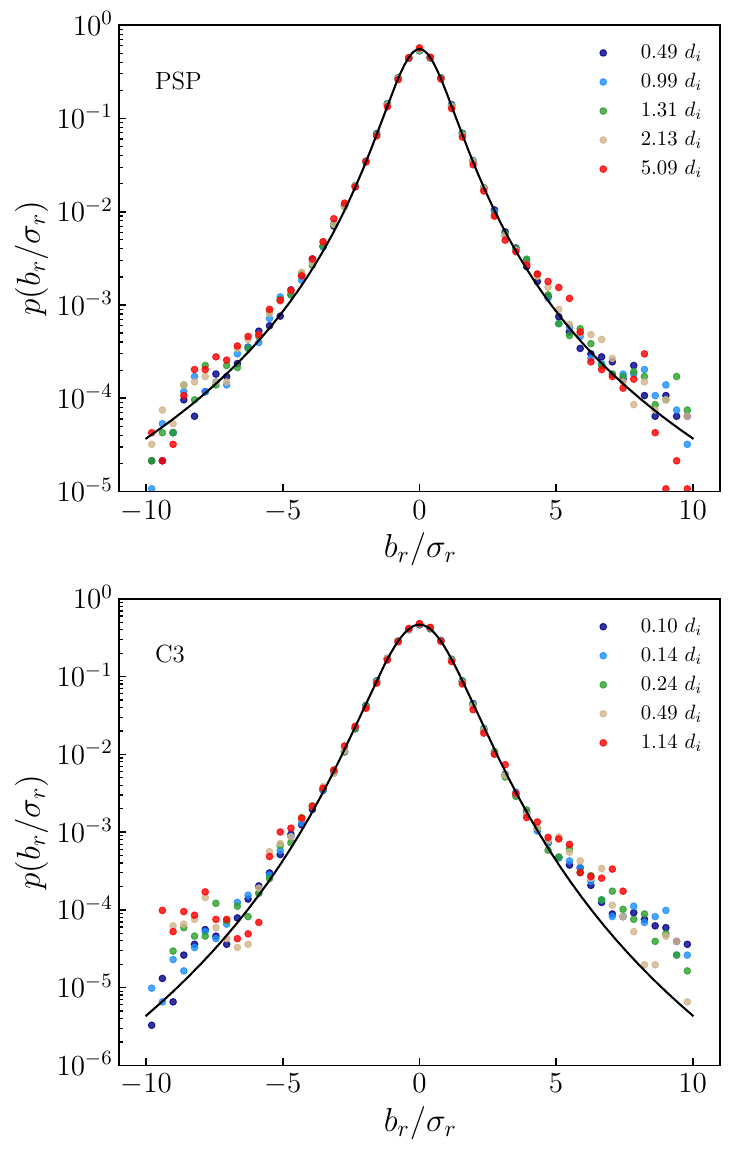}
    \caption{Magnetic field fluctuation PDFs. Filled circles represent the statistics computed on PSP and C3 measurements at different scales, covering about one order of magnitude in the scale separation $r$. Theoretical predictions provided by the Kappa distribution are indicated by solid lines.}
    \label{fig:kappa}
\end{figure}

The data used in this work are reported in Figure \ref{fig:data}. To infer information about the spatial increment vector $\bm{b}=\bm{B}(\bm{x}+\bm{r})-\bm{B}(\bm{x})$ we assume the Taylor's hypothesis to be valid. This hypothesis requires that the Alfvén velocity, which represents the fastest propagating perturbation in plasmas, must be smaller than the flow speed. Average values of the Alfvén velocity are reported in Table \ref{tab:1}. If $\bm{V}$ is the average flow speed, the time-scale separation $\tau$ can be converted in the spatial separation vector through the relation $\bm{r}=\bm{V}\tau$, that fix the sampling direction parallel to the solar wind velocity. Since the solar wind is almost radially expanding, we select the radial component of both magnetic field data samples, i.e., $B_R$ for PSP and the $x$ component in the Geocentric Solar Ecliptic (GSE) reference frame for C3, as representative of the sampling direction, being C3 at the Earth orbit.
The power spectral densities of the two data sample used in the analysis are presented in Figure \ref{fig:psd}. Typical spectral slopes of fully developed turbulence, i.e., Kolmogorov spectrum $f^{-1.7}$, and kinetic scale fluctuations, with steeper trend $f^{-\alpha}$, with $\alpha\in[2.5,2.7]$, are reported for reference. The spectral frequency associated with the smallest scale used in the analysis is indicated by the blue vertical line, corresponding to $\tau=0.020$ s for PSP and $\tau=0.027$ s for C3. The Doppler shifted frequencies associated with ion inertial length $f_d=V/(2\pi d_i)$ and ion gyroradius $f_\rho=V/(2\pi\rho_i)$ are also reported in the figure. By virtute of Taylor's hypothesis, temporal scales can be interpreted as spatial scales and vice versa.

Magnetic field fluctuations developed by turbulence and kinetic-scale dynamics in the solar wind are inherently anisotropic \citep{Horbury2012}. The presence of a strong outward/inward mean field generated by the Sun sets a preferential direction in space which tends to inhibit strong parallel fluctuations. As a result, magnetic field fluctuations develop more intensely in the direction perpendicular to the magnetic field. We define the local mean magnetic field as \begin{equation}
    \bm{B}_{loc}(s, t) = A_s \sum_{k=1}^{N} \bm{B}(t_k)\exp{\left[\frac{-(t_k - t)^2}{2s^2}\right]} 
\end{equation}
where $s$ indicates the scale and $A_s$ the normalization constant 
\begin{equation}
    A_s^{-1} = \sum_{k=1}^{N}
\exp{\left[\frac{-(t_k - t)^2}{2s^2}\right]},
\end{equation}
and we estimate the angle $\theta_{VB}(\bm{x},\bm{r})$ with respect to the sampling direction \citep{Horbury2008,Podesta2009,Podesta2011}. We emphasize that this particular choice for defining the mean local magnetic field does not affect the results. In fact, we carefully verified that the use of different prescriptions, e.g., the moving average of the magnetic field vector with a window of length $10^2d_i$, does not affect the results significantly (not shown). In this work, we consider only fluctuations perpendicular to $\bm{B}_{loc}$ by introducing the condition $80^\circ<\theta_{VB}<100^\circ$ in the magnetic field increments $\bm{b}$. The histograms of the angles $\theta_{VB}$ observed in both PSP and C3 data samples are shown in Figure \ref{fig:theta}. We note that the probability density associated with perpendicular fluctuations, i.e., between the vertical lines, in the case of C3 is slightly higher with respect to the PSP data interval. The scale at which the statistics is displayed is $\tau=0.03$ s, but it is representative of the overall angle statistics due to the weak scale dependence of the probability density shown in Figure \ref{fig:theta} (not shown).

The first step in the analysis is to estimate the scale-by-scale parameters $\kappa$ and $\lambda$ defined in Equation \eqref{eq:kappalam}. These equations rely on the parameterization of the function $D$, which is assumed to be a quadratic polynomial $Q(b_r,r)$. This assumption, justified in Section \ref{sec:phe} as a rather general choice in order to describe turbulent fluctuations in terms of an advection/diffusion dynamics, constitutes a well documented empirical evidence in the small scales of the solar wind \cite{benella2022markovian,benella2022kramers}. By using magnetic field observations we estimate all the scale-dependent parameters $\{Q_0,Q_1,Q_2\}$ appearing in Equations \eqref{Eq:L} and \eqref{Eq:Q} by fitting the Kramers-Moyal (KM) coefficients $D(b_r,r)$ scale-by-scale. This iterative procedure provides the entire set of functions $D_{\Delta}(b_r,r)$ representing the finite-scale approximations of the true second-order KM coefficient $D(b_r,r)$, where $\Delta$ indicates the small step used in the calculation. Therefore, when we use Equation \eqref{Eq:q} in order to link the set of parameters $\{Q_0,Q_1,Q_2\}$ to the parameters $\{q_0,q_1,q_2\}$ of the standardized fluctuation $b_r$, we must take into account the correction terms arising from the finite-scale approximation \citep{benella2022kramers}. Further details on the correct estimation of finite-scale KM coefficients are reported in Appendix \ref{app:b}.

The scale-dependent behavior of the parameters $\lambda$ and $\kappa$ of the generalized distribution \eqref{Eq:genK} is depicted in Figure \ref{fig:kl}. The vertical lines indicate the ion-inertial length as a reference. Whereas large scale data points exhibit large uncertainty in their estimation, due to the limited statistics which rely on subsets of few-hour data samples conditioned to the value of $\theta_{VB}$, the trend of the $\kappa$ and $\lambda$ parameters can be precisely resolved. We observe that the parameter $\lambda$ tends to values close to zero at kinetic scales in both data samples. This implies that the general solution \eqref{Eq:genK} can be approximated by the Kappa distribution reported in Equation \eqref{Eq:kappa}. The parameter $\kappa$ shows a tendency towards stable positive values, which differs in the two considered data sets. PSP fluctuations are associated with a smaller value of $\kappa$, which is about $1.33$, whereas for C3 data we find $\kappa=2.61$.

Equation \eqref{Eq:kappa} provides a prediction for the PDF of magnetic field fluctuations at kinetic scales. The small-scale values assumed by $\kappa$ in PSP and C3 data samples are here used to draw this one-parameter distribution. A comparison to spacecraft observations is displayed in Figure \ref{fig:kappa}. Starting from the smallest scales considered in the analysis (corresponding to the blue vertical lines in Figure \ref{fig:psd}) we show the superposition of about one order of magnitude in scale separation, reaching scales of the order of $d_i$ (circles). Solid lines are the kappa-distribution with parameters $\kappa$ equal to 1.33 for PSP and 2.61 for C3. The interval of perpendicular fluctuation magnitude reported in the figure is $\pm10\sigma_r$. The agreement between theoretical predictions and observation is remarkable especially if we inspect the interval within $\pm5\sigma_r$. Due to the finite size of the samples, in fact, the heavy tails of the distributions are not well sampled, resulting in slightly scattered points and/or cutoffs. However, from a physical point of view a cutoff with respect to the ideal Kappa distribution must be observed as a consequence of the finite energy contained in the fluctuations, thus guaranteeing the finiteness of all high-order moments.

The remarkable agreement between model prediction and data reported in Figure \ref{fig:kappa} is a peculiar property of space-plasma kinetic fluctuations. At larger scales, the turbulent cascade generates an inhomogeneous energy transfer which is associated with the intermittency of fluctuations, in a statistical sense. Intermittency is usually quantified by investigating the scale dependence of the flatness \citep{frisch1995,biskamp_2003}. Changing values of the flatness can be associated with a scale-by-scale modification of the shape of the distribution function of the fluctuations, resulting in the impossibility of obtaining a master curve for the process by rescaling the stochastic variable with only one parameter \citep{Sorriso1999}. Moreover, if we interpret the distribution (\ref{Eq:kappa}) as an instantaneous stationary distribution at a fixed $r$, i.e., by assuming the stationary solution of the FPE to be locally valid, the obtained stationary distributions do not capture the onset of the typical intermittent turbulent dynamics, with a tendency towards a dramatic underestimation of extreme fluctuations \citep{Nickelsen2013}.

\section{Discussion and conclusions}
\label{sec:con}

%
The multi-scale approach developed by \citep{friedrich1997} is effective in catching and reproducing the statistical properties of turbulent fluctuations evolving across scales. This approach, previously extended to space plasma turbulence by \cite{strumik2008testing}, is here specialized to solar wind kinetic scales, where the statistics of magnetic field fluctuations is known to be globally self-similar \citep{kiyani2009global,osman2015}. Therefore, starting from the hypothesis of self-similar scaling, we derived the master curve describing the distribution of magnetic field fluctuations, which is found to be a Kappa distribution. Efforts to link Kappa distributions observed in the statistics of the magnetic field fluctuations to the scale-to-scale stochastic process 
have recently been made in \cite{benella2022markovian}. 
Later on, other authors employed this framework to test the scale invariance of kinetic fluctuations observed in the magnetosheath and close to the magnetopause \citep{macek2022mms,Macek2023MNRAS,Wojcik2024}.
The present work represents a step forward in this field, providing an analytical derivation of the master curve and linking the parameter $\kappa$ of the Kappa distribution to the diffusion term $D(b_r,r)$ associated with interplanetary magnetic field observations. The master curve derived in this work can be either asymmetric with respect to zero, with the asymmetry controlled by the parameter $\lambda$, thus leading to a generalization of the standard Kappa distribution. 

We excluded parallel fluctuations from statistics by selecting magnetic field increments within the $80^\circ<\theta_{VB}<100^\circ$ range. The reason for that is two-fold. First, magnetic field fluctuations are strongly anisotropic and develop mostly in the perpendicular direction with respect to the local mean field, thus being representative of the global statistics. In this case, the master curve, obtained by introducing phenomenological arguments in the framework of stochastic processes, is in agreement with observations. The second reason is that parallel fluctuations encompass ion-resonant processes potentially playing a key role in mediating the collisionless dissipation of turbulent energy \citep{Bowen2024}. Whereas fluctuations associated with dominant circular polarized ion waves exhibit a global scale invariance, data samples having lesser power in the ion-resonant component may develop intermittency. In fact, the absence of circular ion waves is correlated with the occurrence of non-Gaussian intermittent fluctuations, indicative of the presence of small-scale current sheets. Hence, the hypothesis of stationary FPE, on which this work is based, could not always been satisfied for parallel fluctuations.
From the physical side, the existence of a master curve can be related to the presence of homogeneous and globally self-similar fluctuations. In fact, the anomalous scaling observed in turbulence is associated with the inhomogeneity of the energy transfer. Therefore, the regularization of the scaling laws observed in the kinetic range and the emergence of a master curve suggest that energy transfer undergoes an homogenization at small scales. This scenario is in agreement with recent numerical results based on weak kinetic-Alfvén wave turbulence \citep{David2019,David2024}, but also with observations \citep{Richard2024} of energy transfer/dissipation in near-Earth space plasmas.

The parameter $\kappa$ of the master curve defined in Equation \eqref{eq:kappalam} measures the weight of $q_0(r)$ with respect to $q_2(r)$. In other words, $\kappa$ represents the ratio between additive and multiplicative noise in the diffusion term \eqref{eq:diffnorm} of the stochastic process $b_r$. The case of purely additive noise, viz., $q_2(r)\to0$, corresponds to the limit $\kappa\to\infty$ where the master curve \eqref{Eq:kappa} tends towards a Gaussian density function. The heavy tails observed in the magnetic field statistics are thus associated with the multiplicative part of the stochastic dynamics, which plays a fundamental role in the stochastic modeling of kinetic scale fluctuations. This finding helps in pushing forward this description, with the aim of assessing and understanding the meaning of the different terms appearing in the model. In \cite{Benella2023} it has been shown how the drift is the leading term in describing kinetic scale fluctuations which are globally self-similar in average, thus being a proxy for a homogeneous energy transfer. In that case, the contribution of the diffusion term in determining average/global quantities, such as the scaling exponent, was neglected. In the present work, the study of the probability density function, instead of its moments, allows us to assess that the diffusion term plays a key role in defining shape and asymmetry of the fluctuation PDF.

The derivation of $\kappa$ and $\lambda$ from the Langevin process provides us a pair of parameters easily computed from spacecraft data which fully describe the statistics of kinetic scale fluctuations. A systematic investigation of these parameters as a function of relevant quantities, such as radial distance from the Sun, Reynolds number, average magnitude of the magnetic field, etc., is here left as a future perspective, with the idea of relating it to a possible universal behavior of the fluctuations at these scales. Finally, we emphasize that the functional form of drift and diffusion terms (empirical evidence) is based on the scaling arguments summarized in Section \ref{sec:phe}. The resulting stochastic process is associated with a master curve which resembles spacecraft in situ observations, thus providing interesting constraints on the phenomenology of small-scale dynamics. Indeed, future theoretical frameworks should include the statistical properties presented in this paper. The search for a phenomenology leading to Equations \eqref{Eq:L} and \eqref{Eq:Q} thus represents the necessary and crucial step to take in future investigations.

\begin{acknowledgements}
Authors acknowledge Emiliya Yordanova for providing Cluster data and for the fruitful discussions. This work is supported by the mini-grant “The solar wind: a paradigm for complex system dynamics" financed by the National Institute for Astrophysics under the call “Fundamental Research 2022". M.S acknowledges the mini-grant “Investigating the Universal Nature of Magnetic field Fluctuations in Solar Wind Turbulence: A Multifractal Approach" financed by the National Institute for Astrophysics under the call “Fundamental Research 2023". PSP, Wind and SO data used in this study are available at the NASA Space Physics Data Facility (SPDF), \url{https://spdf.gsfc.nasa.gov/index.html}. The authors acknowledge the contributions of the FIELDS team to the Parker Solar Probe mission, the Cluster FGM, STAFF P.I.s and teams, and the ESA-Cluster Science Archive for making available the data used in this work.
\end{acknowledgements}

\bibliographystyle{aa}
\bibliography{kappa_refs}

\begin{appendix}
\section{Derivation of the master curve}\label{app:a}
For general purposes and a more clear connection with the literature, we will use here $t\geq0$ as independent “time” variable for the Fokker-Planck description of a real Itô stochastic process $(X_t)_{t\geq0}$ satisfying \begin{equation}
    dX_t=M(X_t,t)dt+\sqrt{2D(X_t,t)}dW_t, \label{Eq:langevin_t}
\end{equation}
with $(W_t)_{t\geq0}$ being a standard Wiener process. All the formulas in the remaining text can be obtained via some suitable change of variable, such as $r=r_0 e^{-t/\tau}$, or $r=r_0-Vt$, together with a redefinition of the drift $M$ and the diffusion coefficient $D$. Notice that $r$ decreases while $t$ increases.

The probability density $\rho_t(X)$ of finding $X_t$ in the state $X$ is defined by the relation
\begin{equation}
    \mathbb{P}(X_t\in I) = \int_I \rho_t(X) dX
\end{equation}
for any subset $I\subseteq\mathbb{R}$ and it has a unitary integral over the definition domain
\begin{equation}
    \int_{\mathbb{R}} \rho_t(X) dX = 1. \label{Eq:probability_conservation}
\end{equation}
Assuming the initial density function $\rho_0(X)$ to be known, the generalized Langevin equation (\ref{Eq:langevin_t}) is equivalent to the FPE \citep{risken1996fokker}
\begin{equation}
    \frac{\partial\rho_r(X)}{\partial t} = \frac{\partial}{\partial X}\biggl[\frac{\partial}{\partial X}(D(X,t)\rho_t(X)) - M(X,t)\rho_t(X)\biggr].
\end{equation}
The derivation of Equation (\ref{Eq:probability_conservation}) with respect to $t$ gives the following condition on the boundary terms:
\begin{equation}%
    \biggl[\frac{\partial }{\partial X}(D(X,t)\rho_t(X)) - M(X,t)\rho_t(X)\biggr]^{+\infty}_{-\infty} = 0. \label{Eq:boundary_rerms_condition}
\end{equation}%
We define the mean and the variance of $X_t$ as
\begin{align}
    &\mu_t := \int_{\mathbb{R}} X\rho_t(X)dX, \\ &\sigma^2_t := \int_{\mathbb{R}} X^2\rho_t(X)dX - \mu_t^2,
\end{align}
respectively, and we assume that both are finite. By deriving $\mu_t$ and $\sigma^2_t$ with respect to $t$, we obtain
\begin{equation}%
    \frac{d\mu_t}{dt} = \int_{\mathbb{R}} M(X,t)\rho_t(X)dX - \biggl[D(X,t)\rho_t(X)\biggr]^{+\infty}_{-\infty} \label{Eq:dmu/dt}
\end{equation}%
and
\begin{multline}
    \frac{d\sigma^2_t}{dt} = 2\int_{\mathbb{R}} (D(X,t) + XM(X,t))\rho_t(X)dX - 2\mu_t\frac{d\mu_t}{dt}\\ - 2\biggl[XD(X,t)\rho_t(X)\biggr]^{+\infty}_{-\infty}, \label{Eq:dsigma/dt}
\end{multline}%
where Equation (\ref{Eq:boundary_rerms_condition}) has been used. 
Assuming the polynomial drift and diffusion terms introduced in Equations (\ref{Eq:L}) and (\ref{Eq:Q}) and assuming vanishing boundary terms in Equations (\ref{Eq:dmu/dt}, \ref{Eq:dsigma/dt}), we get the following closed system of differential equations
\begin{equation}%
    \frac{d\mu_t}{dt} = L_1(t)\mu_t + L_0(t),\label{Eq:dmu/dt_LQ}
\end{equation}%
\begin{equation}%
    \frac{d\sigma^2_t}{dt} = 2[L_1(t) + Q_2(t)]\sigma^2_t + 2Q(\mu_t,t), \label{Eq:dsigma2/dt_LQ}
\end{equation}%
whose solutions are
\begin{multline}
    \mu_t = \mu_0\exp\biggl(\int_0^tL_1(u)du\biggr)\\ + \int_0^t L_0(s)\exp\biggl(\int_s^t L_1(u)du\biggr)ds,
\end{multline}
and
\begin{multline}
    \sigma^2_t = \sigma^2_0\exp\biggl(2\int_0^t[L_1(u) + Q_2(u)]du\biggr)\\ + 2\int_0^tQ(\mu_s,s)\exp\biggl(2\int_s^t[L_1(u) + Q_2(u)]du\biggr)ds.
\end{multline}
By introducing the standardized process $(x_t)_{t\geq0}$
\begin{equation}
    x_t := \frac{X_t - \mu_t}{\sigma_t},
\end{equation}
the probability density $p_t(x)$ of finding $x_t$ at $x$ is given by
\begin{equation}
    p_t(x) = \sigma_t\rho_t(\mu_t+\sigma_tx).
\end{equation}
It satisfies the following FPE
\begin{multline}
    \frac{\partial p_t(x)}{\partial t} = \frac{\partial }{\partial x}\biggl[ \frac{\partial }{\partial x}(d(x,t)p_t(x)) - m(x,t)p_t(x)\biggr],
    \label{Eq:FPE}
\end{multline}
with drift and diffusion coefficients given by
\begin{align}
    &m(x,t) := \frac{M(\mu_t+\sigma_t x,t) - \frac{d\mu_t}{dt} - \frac{d\sigma_t}{dt}x}{\sigma_t}, \\&d(x,t) := \frac{D(\mu_t+\sigma_t x,t)}{\sigma_t^2}.
\end{align}
In the special case of $M\equiv L$ and $D\equiv Q$, e.g., see Equations \eqref{Eq:L} and \eqref{Eq:Q}, we get $m\equiv l$ and $d\equiv q$, where
\begin{align}%
    &l(x,t) := - [q_0(t) + q_2(t)]x,\\ &q(x,t) := q_0(t) + q_1(t)x + q_2(t)x^2, \label{Eq:lq}
\end{align}%
where the terms $q_0$, $q_1$ and $q_2$ are
\begin{align}%
    &q_0(t) := \frac{Q(\mu_t,t)}{\sigma_t^2}, \\&q_1(t) := \frac{Q_1(t) + 2Q_2(t)\mu_t}{\sigma_t}, \\&q_2(t) := Q_2(t).
\end{align}%
If the process $(x_t)_{t\geq0}$ is stationary, viz., $\partial p_t/\partial t\equiv0$, we can write $p_t(x)\equiv p(x)$ and hence
\begin{equation}%
    \frac{\partial}{\partial x}(d(x,t)p(x)) - m(x,t)p(x) = f(t). \label{Eq:SFP=f}
\end{equation}%
Furthermore, $f(t)$ must vanish identically, as a consequence of \eqref{Eq:dmu/dt}. The solution can then be written as
\begin{equation}%
    p(x) = p(\bar{x})\exp\biggl[\int_{\bar{x}}^x\frac{m(y,t) - \partial d(y,t)/\partial y}{d(y,t)} dy\biggr], \label{Eq:p(x)_general}
\end{equation}%
with some arbitrary $\bar{x}$, which can be suitably chosen. The probability density $p$ is called a master curve.

The explicit dependence on $t$ in \eqref{Eq:p(x)_general} must be completely reabsorbed, for the probability density to be stationary, thus we obtain the condition
\begin{equation}%
    \frac{m(x,t) - \partial d(x,t)/\partial x}{d(x,t)} = C(x). \label{Eq:master_curve_condition}
\end{equation}%
In the special case when $m\equiv l$ and $d\equiv q$, we get $C\equiv C_{\kappa,\lambda}$, with
\begin{equation}
    C_{\kappa,\lambda}(x) := \frac{2[\lambda - (\kappa+1)x]}{2\kappa - 1 - 2\lambda x + x^2},
\end{equation}
where $2\kappa-1>\lambda^2$, leading to
\begin{align}
    &\lambda = - \frac{q_1(t)}{2q_2(t)}, \\&\kappa = \frac{1}{2} + \frac{q_0(t)}{2q_2(t)}.
\end{align}
In this case, by choosing $\bar{x}=\lambda$, the corresponding stationary probability density follows to be
\begin{multline}
    p_{\kappa,\lambda}(x) = N_{\kappa,\lambda}f_{\kappa}\biggl(\tfrac{x - \lambda}{\sqrt{2\kappa - 1 - \lambda^2}}\biggr)\exp\biggl(- \tfrac{2\lambda\kappa}{\sqrt{2\kappa - 1 - \lambda^2}}\tan^{-1}\biggl(\tfrac{x - \lambda}{\sqrt{2\kappa - 1 - \lambda^2}}\biggr)\biggr),
\end{multline}
where
\begin{align}
    &f_{\kappa}(\xi) := (1 + \xi^2)^{-\kappa-1},
    \\&N_{\kappa,\lambda} := \frac{2^{2\kappa}\Gamma\left(\kappa + 1 + i\frac{\lambda\kappa}{\sqrt{2\kappa - 1 - \lambda^2}}\right)\Gamma\left(\kappa + 1 - i\frac{\lambda\kappa}{\sqrt{2\kappa - 1 - \lambda^2}}\right)}{\pi\sqrt{2\kappa - 1 - \lambda^2}\Gamma(2\kappa + 1)},
\end{align}
with $i$ being the imaginary unit and $\Gamma$ the Euler gamma function:
\begin{equation}
    \Gamma(z) = \int_{0}^{\infty} u^{z-1} e^{-u} du.
\end{equation}
To get the normalization constant $N_{\kappa,\lambda}$ we made use of the following identity
\begin{equation}\label{Eq:integral}
    \int_{-\infty}^{+\infty} f_{\kappa}(\xi) e^{2a\tan^{-1}\xi}d\xi = \frac{\pi\kappa\Gamma(2\kappa)}{2^{2\kappa-1}\Gamma(\kappa + 1 + ia)\Gamma(\kappa + 1 - ia)},
\end{equation}
together with
\begin{multline}
    B_{\frac{1}{2}}(i a-\kappa ,-i a-\kappa )+B_{\frac{1}{2}}(-i a-\kappa
   ,i a-\kappa )
   \\= \frac{4\pi\kappa \sin (2 \pi \kappa) \Gamma (2 \kappa
   )}{(\cos (2 \pi  \kappa
   )-\cosh (2 \pi  a))\Gamma (\kappa +1+i a)\Gamma (\kappa +1-i a)},
\end{multline}
which has been found as a condition for the integral (\ref{Eq:integral}) to be real, where the incomplete beta function is defined as
\begin{equation}
    B_{z}(a,b) = \int_{0}^{z} u^{a-1} (1-u)^{b-1} du.
\end{equation}

\section{Finite-scale estimation of KM coefficients}\label{app:b}
\begin{figure}
    \centering
    \includegraphics[width=0.5\textwidth]{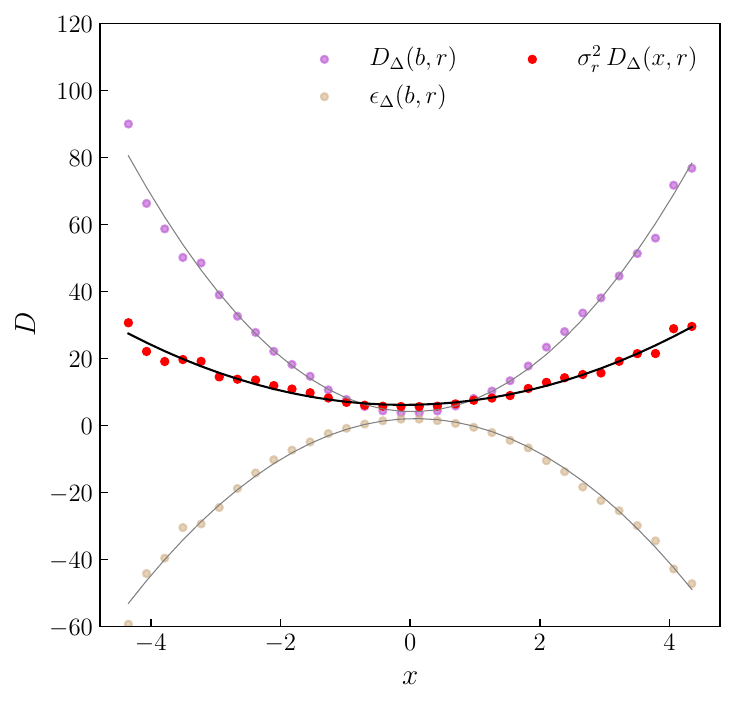}
    \caption{Example of finite scale KM coefficient estimation. Filled circles represent non parametrical estimation of $D_\Delta(b,r)$, purple, $\epsilon_\Delta(b,r)$, beige, and $\sigma_r^2\,D_\Delta(x,r)$, red. Solid lines indicate the second-order polynomial fits.}
    \label{fig:d2_corr}
\end{figure}
An interesting feature in the derivation of the drift-diffusion parameters for the standardized variable $x$ is their dependence upon the diffusion coefficient $D$ of the magnetic field fluctuations only. All the relations derived in Section \ref{sec:met} are based on the calculation of the KM coefficients, which are the limit of the conditional moments \citep{risken1996fokker}. Hence, the second-order KM coefficient (viz., diffusion term) can be written as
\begin{equation}
    D(b,r)=\lim_{\Delta\to0}\frac{1}{2!\Delta}\mathbb{E}[(b(r-\Delta)-b(r))^2|b(r)].
\end{equation}
The limit of vanishing separation $\Delta$ is a daunting task when dealing with "real-world" timeseries. In this work we set the separation $\Delta$ to be finite and small, close to the resolution of the data, but large enough to avoid effect due to the instrumental noise at high frequencies. When the values of $\Delta$ is finite, although arbitrary small, we need to consider the correction terms appearing in the change from the variable $b$ to the standardized variable $x=b/\sigma_r$. In the ideal case of vanishing $\Delta$ we used the relation
\begin{equation}
    \sigma_r^2\,D(x,r)=D(b,r).
\end{equation}
According to \cite{benella2022kramers}, in the case of finite and arbitrary small $\Delta$, we must introduce the following correction
\begin{equation}
        \sigma_r^2\,D_\Delta(x,r)=D_\Delta(b,r)+\epsilon_\Delta(b,r),
\end{equation}
with $\epsilon_\Delta(b_r, r)$ defined as \citep{benella2022kramers}
\begin{multline}
    \epsilon_\Delta(b_r,r)= \frac{1}{2\Delta}\biggl(\frac{\sigma_r^2}{\sigma_r'^2}-1\biggr)\,\mathbb{E}[b^2(r-\Delta)|b(r)] \\+\frac{1}{\Delta}\biggl(1-\frac{\sigma_r}{\sigma_r'}\biggr)\,\mathbb{E}[b(r-\Delta)b(r)|b(r)].
\end{multline}
The correction term $\epsilon_\Delta(b_r,r)$ exhibits a quadratic trend as the diffusion term $D_\Delta(b_r,r)$, thus we can introduce a quadratic parametrization also for $\epsilon_\Delta(b_r,r)$.
An graphical example of the transformation between $D_\Delta(b,r)$ and $D_\Delta(x,r)$ calculated on C3 data at the fixed scale $r=0.1\,d_i$ is shown in Figure \ref{fig:d2_corr}. Circles indicate $D_\Delta(b,r=0.1\,d_i)$ in purple, the correction $\epsilon_\Delta(b,r=0.1\,d_i)$ in beige, and the diffusion term of the standardized variable $\sigma_r^2\,D_\Delta(x,r=0.1\,d_i)$ in red, while solid lines show their second-order polynomial fits.

\end{appendix}

\end{document}